\documentclass[twocolumn,prb,amsmath,amssymb,superscriptaddress,showpacs]{revtex4}

\usepackage{graphicx}
\usepackage{dcolumn}
\usepackage{bm}

\usepackage[dvips]{color}

\newcommand {\imag}{{\text{i}}}

\begin{document}

\title{Metamaterial inspired perfect tunneling in semiconductor heterostructures}

\author{L. Jelinek}
\email{l\textunderscore jelinek@us.es} 
\affiliation{Department of Electromagnetic Field, Czech Technical University in Prague, 166 27-Prague, Czech Republic}

\author{J. D. Baena}
\email{jdbaenad@unal.edu.co} 
\affiliation{Departamento de Fisica, Universidad Nacional de Colombia, Bogota, Colombia}

\author{J. Voves}
\email{voves@fel.cvut.cz} 
\affiliation{Department of Microelectronics, Czech Technical University in Prague, 166 27-Prague, Czech Republic}

\author{R. Marques}
\email{marques@us.es} 
\affiliation{Departamento de Electr\'onica y Electromagnetismo, Universidad de Sevilla, 41012-Sevilla, Spain}



\date{\today}

\begin{abstract}
In this paper we are using formal analogy of electromagnetic wave equation and Schr{\"o}dinger equation in order to study the phenomenon of perfect tunneling (tunneling with unitary transmittance) in 1D semiconductor heterostructure. Using the Kane model of semiconductor we are showing that such phenomenon can indeed exist, resembling all the interesting features of the analogous phenomenon in classical electromagnetism in which metamaterials (substances with negative material parameters) are involved. We believe that these results can open up the way to interesting applications in which the metamaterial ideas are transfered into semiconductor domain.
\end{abstract}


\maketitle

\section{Introduction}

The tunneling of electrons through potential barrier is a phenomenon known to physicists for long \cite{Gamow-1928, Gurney-1929}. At first, the tunneling amplitudes were known to take appreciable values only on atomic scales, later on however, the use of resonant tunneling found in semiconductor layered structures \cite{Esaki-1970} opened the way to high tunneling amplitudes even in macroscopic devices, such as resonant tunneling diodes.

Phenomenon equivalent to quantum tunneling is known also in other fields of physics, one example being classical electromagnetism \cite{Jackson01}. There, for example, a section of hollow metallic waveguide above cutoff frequency can serve as the environment where the wave propagates, while the section of waveguide below cutoff frequency can serve as the potential barrier through which the photons can tunnel. It was also the field of electromagnetism where the so called perfect tunneling, which is the tunneling with unitary transmission coefficient, has been proposed \cite{Pendry-2000}, theoretically studied \cite{Alu-2003}, \cite{Smith-2003} and experimentally proven \cite{Baena-2005a} with the help of metamaterials \cite{Marques} which are substances offering at some frequency negative values of permittivity and permeability.

The aim of this paper is to show that perfect tunneling, like other metamaterial inspired phenomena, such as negative refraction \cite{Cheianov-2007}, can exist also in quantum domain. The proposal is based on mathematical similarity of Schr{\"o}dinger equation and electromagnetic wave equation.

\section{Maxwell-Schr{\"o}dinger Analogy}

In order to proceeded to perfect quantum tunneling, let us first show the example of perfect tunneling setup in the case of electromagnetic waves. The structure is sketched in Fig. 1 and is following the idea \cite{Alu-2003}. The layers are assumed laterally infinite and the plane waves are supposed to propagate perpendicularly to the layers. The structure is fed by incident wave from the vacuum region 1, which potentially also contains some reflected wave. Regions 2,3,4 support only evanescent waves due to the negative values of constitutive parameters. Any wave that tunnels the structure will appear in region 5. It has been shown \cite{Alu-2003} that it is always possible to find such $d_1, d_2$ for which the tunneling (transmission) through this structure is equal to unity and can thus be called perfect, even though the structure contains potential barriers of theoretically any thickness. Realistic example of this structure has been experimentally studied \cite{Baena-2005a} with the use of metamaterials. Note also that this perfect tunneling setup is not limited to three layers, but any number of such alternating negative parameter layers can be used \cite{Shamonina-2001, Baena-2005b} leading to the same result.

\begin{figure}
\centering
\includegraphics[width=0.8\columnwidth]{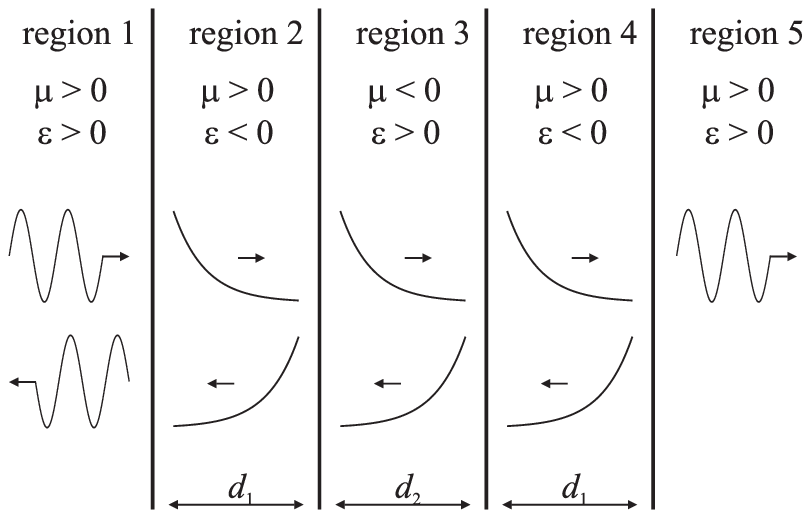}
\caption{\label{Fig1} Sketch of the electromagnetic perfect tunneling setup.}
\end{figure}

The way to transform above structure in a quantum one uses the analogy between electromagnetic wave equation and Schr{\"o}dinger equation \cite{Gaylord-1993,Dragoman-1999,Dragoman-2007}. More specifically, if the longitudinal axis in Fig. 1 is denoted as $z$-axis the Maxwell equations for monochromatic plane wave with angular frequency $\omega$ propagating along this axis can be written as

\begin{equation}
\frac{\partial }
{{\partial z}}\left[ {\begin{array}{*{20}c}
   {E_x }  \\
   {H_y }  \\

 \end{array} } \right] = \left[ {\begin{array}{*{20}c}
   0 & { \imag \omega \mu }  \\
   { \imag \omega \varepsilon } & 0  \\

 \end{array} } \right]\left[ {\begin{array}{*{20}c}
   {E_x }  \\
   {H_y }  \\

\end{array} } \right].
\end{equation}
The Eq. 1 has to be accompanied by proper boundary conditions for each boundary between different material regions, namely
\begin{equation}
E_x^ +   = E_x^ -
\end{equation}
and
\begin{equation}
\frac{1}
{{\mu ^ +  }}\frac{{\partial E_x^ +  }}
{{\partial z}} = \frac{1}
{{\mu ^ -  }}\frac{{\partial E_x^ -  }}
{{\partial z}}.
\end{equation}
On the other hand, in the uni-band approximation, the time independent Schr{\"o}dinger equation for the wavefunction envelope $\psi$ can be written as \cite{Wannier-1937,Slater-1949,BenDaniel-1966}

\begin{equation}
\frac{{ - \hbar ^2 }}
{{2m}}\frac{{\partial ^2 \psi }}
{{\partial ^2 z}} + V\psi  = E\psi,
\end{equation}
with $m$ as the effective mass of the particle in a given material, $E$ as energy of the particle and $V$ as potential step on the boundary between two adjacent materials. Equation (4) can be further rewritten as 
\begin{equation}
\frac{\partial }
{{\partial z}}\left[ {\begin{array}{*{20}c}
   \psi   \\
   {\frac{{-\imag \hbar }}
{m}\frac{{\partial \psi }}
{{\partial z}}}  \\

 \end{array} } \right] = \left[ {\begin{array}{*{20}c}
   0 & { \imag \frac{m}
{\hbar }}  \\
   { 2 \imag \frac{{\left( {E - V} \right)}}
{\hbar }} & 0  \\

 \end{array} } \right]\left[ {\begin{array}{*{20}c}
   \psi   \\
   {\frac{{-\imag \hbar }}
{m}\frac{{\partial \psi }}
{{\partial z}}}  \\

 \end{array} } \right],
\end{equation}
and accompanied by proper boundary conditions, namely
\begin{equation}
\psi^ +   = \psi^ -
\end{equation}
and
\begin{equation}
\frac{1}
{{m ^ +  }}\frac{{\partial \psi^ +  }}
{{\partial z}} = \frac{1}
{{m ^ -  }}\frac{{\partial \psi^ -  }}
{{\partial z}}.
\end{equation}
Note that Eq. 7 has replaced here the usual condition for continuity of the derivative. This condition arise from the electric current conservation on the material boundary \cite{Harrison-1961}, where the quantum analog of electric current density is taken as ${\mathbf{j}} = e {\mathbf{s}}$, with $e$ as the electron charge and
\begin{equation}
{\mathbf{s}} =  \frac{{\hbar }}
{{2 \imag m}}\left( {\psi ^* \nabla \psi  - \psi \nabla \psi ^* } \right)
\end{equation}
as the probability density current.

Comparing (1,2,3) with (5,6,7) one can see that the Maxwell-Schr{\"o}dinger analogy can be made perfect and that the solution of (1) on the structure of Fig. 1 will be mathematicaly identical with the quantum structure where
\begin{equation}
\begin{gathered}
  E_x  \to \psi  \hfill \\
  \mu  \to m \hfill \\
  \varepsilon  \to 2\left( {E - V} \right) \hfill \\
  \omega  \to 1/\hbar.  \hfill \\
\end{gathered} 
\end{equation}
It is also interesting to note that (8) is giving another part of the Maxwell-Schr{\"o}dinger analogy, namely the analogy of the Pointing vector and probability density current
\begin{equation}
\begin{gathered}
  \operatorname{Re} \left[ {E_x H_y^* } \right] \to s_z. \hfill \\
\end{gathered} 
\end{equation}

\section{Theoretical Analysis}

We are thus about to study the structure shown in Fig. 2. 
\begin{figure}
\centering
\includegraphics[width=0.8\columnwidth]{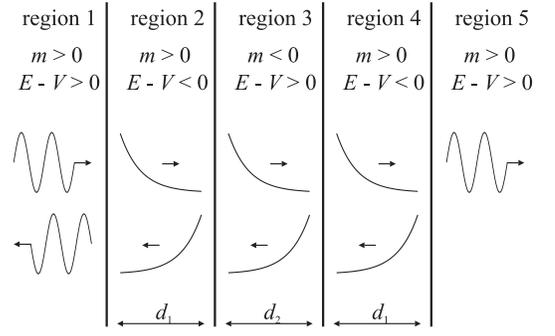}
\caption{\label{Fig2} Sketch of the quantum perfect tunneling setup.}
\end{figure}
Unfortunately, being rigorous, the equation (4) is not valid in any realistic semiconductor heterostructure. Truly the heterostructure of two materials A and B should be better described by 8x8 Kane \cite{Kane-1957, Bastard} model 
\begin{equation}
\begin{gathered}
  \left[ {E _{l0}  - \frac{{\hbar ^2 }}
{{2m_0 }}\frac{{\partial ^2 }}
{{\partial z^2 }}} \right]f_l \left( z \right) +  \hfill \\
   + \sum\limits_{m = 1}^8 {\frac{1}
{{m_0 }}} \left\langle {u_{l0} } \right|\frac{\hbar }
{i}\frac{\partial }
{{\partial z}}\left| {u_{m0} } \right\rangle \frac{\hbar }
{i}\frac{{\partial f_m \left( z \right)}}
{{\partial z}} = E f_l \left( z \right) \hfill \\ 
\end{gathered} 
\end{equation}
with $l=1..8$, with $E _{l0}$ as the double degenerated (spin) band edge energies (for $k=0$) of the conduction, light hole valence, heavy hole valence and split off valence bands ($E _{l0}$ is different for A and B material), with $f_l \left( z \right)$ as the corresponding envelope functions, with $m_0$ as electron mass and with $u_{l0}$ as the bulk material eigenfunctions for the mentioned bands at $k=0$ which are assumed to be the same in the entire heterostructure. In the following we will denote the eight band edge states by their symmetry properties. They will be: $\Gamma _6$ as s-like state with eigenvalues of angular momentum $J=1/2, J_z = \pm 1/2$, $\Gamma _8$ as p-like state with $J=3/2, J_z = \pm 1/2, \pm 3/2$ and $\Gamma _7$ as p-like state with $J=1/2, J_z = \pm 1/2$. 

The matrix system (11) represents the multiband nature of the semiconductors, phenomenon that is not present in our electromagnetic problem. However, it was shown \cite{Bastard-1982, Bastard} that the heavy hole states are uncoupled to the three light states (electron, light hole, split off hole) and that (11) can be directly separated into one heavy hole scalar equation and 3x3 matrix system for light states. Furthermore, under very reasonable approximation \cite{Bastard-1982} (valid for most of the III-V and II-VI compound semiconductors) the 3x3 light states matrix system can be solved for conduction states leading to scalar equation 
\begin{equation}
\left[ { - \frac{{\hbar ^2 }}
{{2m}}\frac{{\partial ^2 }}
{{\partial z^2 }} + E_{\Gamma _6 } } \right]f_{\text{c}} \left( z \right) = E f_{\text{c}} \left( z \right)
\end{equation}
where
\begin{equation}
\frac{1}
{m} = \frac{{2P^2 }}
{3}\left[ {\frac{2}
{{E - E_{\Gamma _8 } }} + \frac{1}
{{E - E_{\Gamma _7 } }}} \right]
\end{equation}
is energy and position dependent effective mass. The quantities $E_{\Gamma _6 } ,E_{\Gamma _8 }, E_{\Gamma _7 } $
are the band edge energies of $\Gamma _6 ,\Gamma _8 , \Gamma _7$ bands which are position dependent in step like manner along the heterostructure. The quantity $P = \frac{{ - i}}{{m_0 }}\left\langle S \right|\frac{\hbar }{i}\frac{\partial }{{\partial x}}\left| X \right\rangle = \frac{{ - i}}{{m_0 }}\left\langle S \right|\frac{\hbar }{i}\frac{\partial }{{\partial y}}\left| Y \right\rangle = \frac{{ - i}}{{m_0 }}\left\langle S \right|\frac{\hbar }{i}\frac{\partial }{{\partial z}}\left| Z \right\rangle $ is the element of the Kane matrix, where $\left| S \right\rangle, \left| X \right\rangle, \left| Y \right\rangle, \left| Z \right\rangle$ represent the s-like and three p-like eigenfunctions for $k=0$.

The envelope equation (12) is accompanied by appropriate boundary conditions that read
\begin{equation}
f_{\text{c}} \left( {z^ +  } \right) = f_{\text{c}} \left( {z^ -  } \right)
\end{equation}
and
\begin{equation}
\frac{1}
{{m^ +  }}\frac{{\partial f_{\text{c}} \left( {z^ +  } \right)}}
{{\partial z}} = \frac{1}
{{m^ -  }}\frac{{\partial f_{\text{c}} \left( {z^ -  } \right)}}
{{\partial z}}
\end{equation}
Thus, after all, the equation for $f_{\text{c}} \left( z \right)$ and its boundary conditions are analogous to
equations (1-3) with the only difference that the effective mass is now function of energy and some adjustable parameters. 


\section{Practical Implementation}

Comparing the scheme of Fig. 2 with mathematical formalism of Sec. III we can arrive to the possible realistic implementation of the perfect tunneling setup using ${\text{Hg}}_{1 - x} {\text{Cd}}_x {\text{Te}}$ ternary alloy. Heterostructures of this alloy were extensively studied in past \cite{Guldner-1983,Chang-1985,LinLiu-1985} for the possibility of the existence of the interface states which are indeed closely related to the perfect tunneling proposed in this paper. The parameters needed for the calculation of the envelope function via (12) and (13) can be obtained from reliable measurements or first principle calculations \cite{Kowalczyk-1986,Johnson-1988,Mecabih-2000} which suggest $2m_0 P^2  \approx 18.5\;{\text{eV}}$, $E_{\Gamma _6 }  \approx \left( {1.47x + 0.08} \right)\;{\text{eV}}$, $E_{\Gamma _8 }  \approx \left( { - 0.36x + 0.36} \right)\;{\text{eV}}$ and $E_{\Gamma _7 }  \approx \left( { - 0.36x - 0.59} \right)\;{\text{eV}}$,where $x$ represents ${\text{Hg}}_{1 - x} {\text{Cd}}_x {\text{Te}}$ mole fraction. These values are also in agreement with recent review \cite{Rogalski-2005} of the HgCdTe alloys.

The possibilities of the ${\text{Hg}}_{1 - x} {\text{Cd}}_x {\text{Te}}$ alloy are graphically represented in Fig. 3. The proposed perfect tunneling demands three different combination of $m$ and $(E - V)$ at a given value of energy, condition which is satisfied for any energy between the two dotted lines in Fig. 3. The obvious possibility is thus the setup depicted in Fig. 4 which also represents the setup with highest energy band of operation allowing thus looser choice of structural dimensions.

\begin{figure}
\centering
\includegraphics[width=0.8\columnwidth]{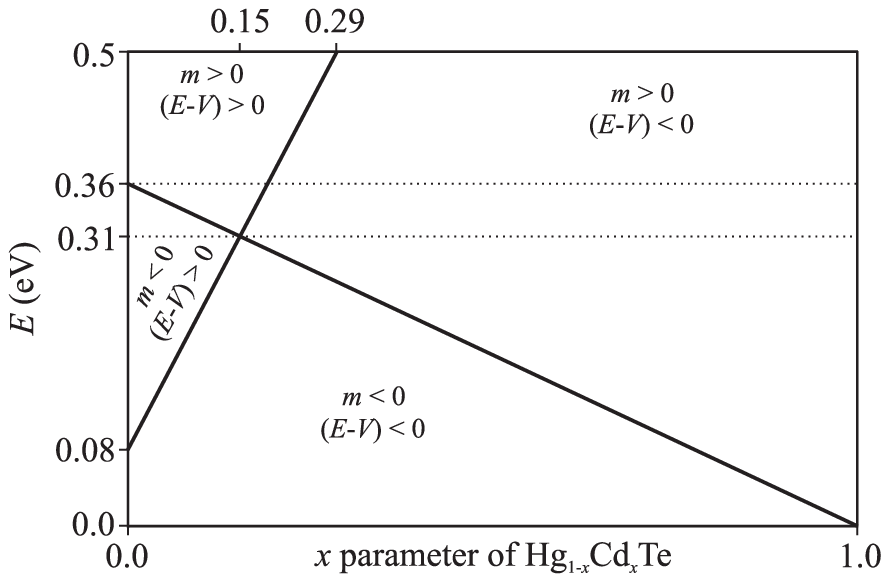}
\caption{\label{Fig3} E-x plane of ${\text{Hg}}_{1 - x} {\text{Cd}}_x {\text{Te}}$ ternary alloy. Solid lines are dividing the plane in four fields with different signs of mass and energy difference. Dotted lines define the band of energies of possible perfect tunneling.}
\end{figure}

\begin{figure}
\centering
\includegraphics[width=0.8\columnwidth]{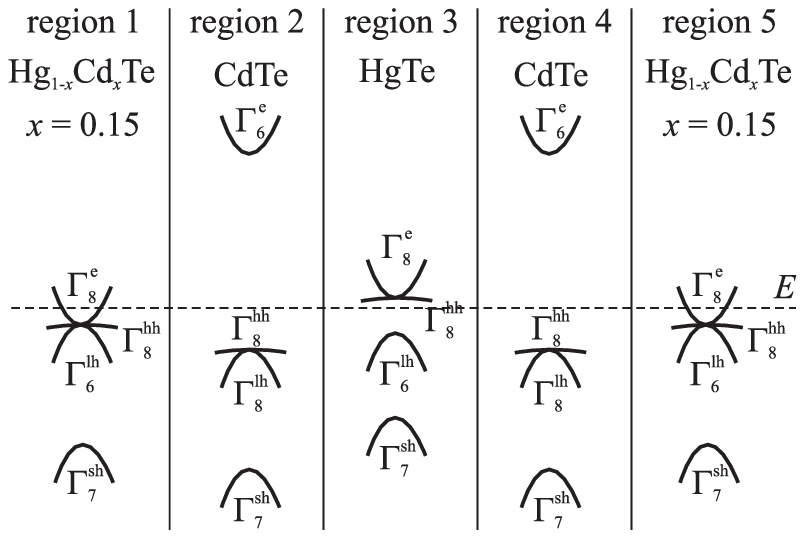}
\caption{\label{Fig4} Band diagram sketch of the realistic quantum tunneling structure.}
\end{figure}

The transmission coefficient in the perfect tunneling energy band and the amplitude of envelope function at the transmission maximum were calculated for $d_1 = 1.26\; \text{nm}$, $d_2 = 10\; \text{nm}$ and are depicted in Fig. 5 and Fig. 6. Both figures clearly present the perfect tunneling we are looking for, including the interface states maxima \cite{Guldner-1983,Chang-1985,LinLiu-1985} at the boundaries where the effective mass change its sign. Analogously to the electromagnetic problem it can be also shown that at the energy of unitary transmittance the probability density current (8) will be constant along all the heterostructure. At this point it is important to stress that the unitary transmission is achieved at energies for which the regions 2,3,4 support only evanescent waves. Such phenomenon is thus quite different from the usual resonant tunneling for which the region 3 is propagative with either $m > 0, \left(E - V \right) > 0$ (resonant tunneling diode) or $m < 0, \left(E - V \right) < 0$ (interband resonant tunneling diode).

\begin{figure}
\centering
\includegraphics[width=0.8\columnwidth]{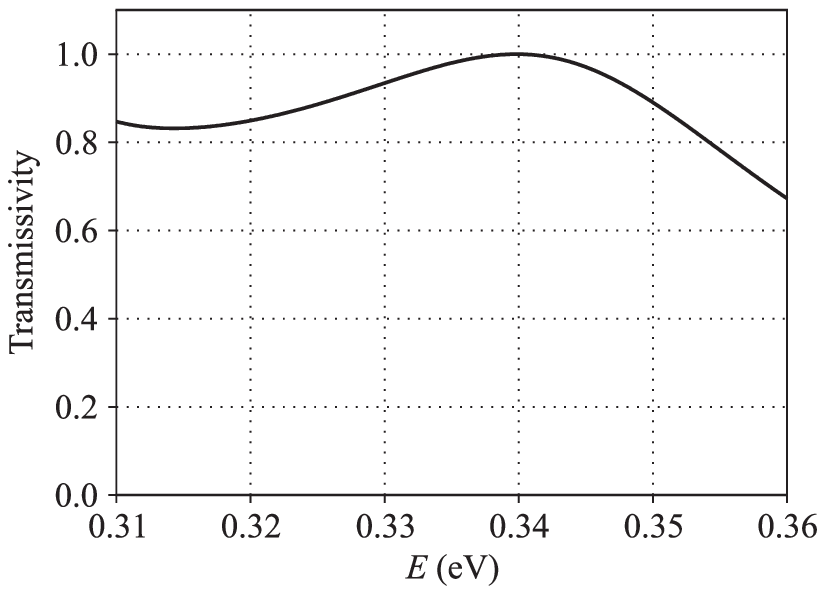}
\caption{\label{Fig5} Amplitude of the transmission coefficient through the structure of Fig. 4 for $d_1 = 1.26\; \text{nm}$, $d_2 = 10\; \text{nm}$.}
\end{figure}
\begin{figure}
\centering
\includegraphics[width=0.8\columnwidth]{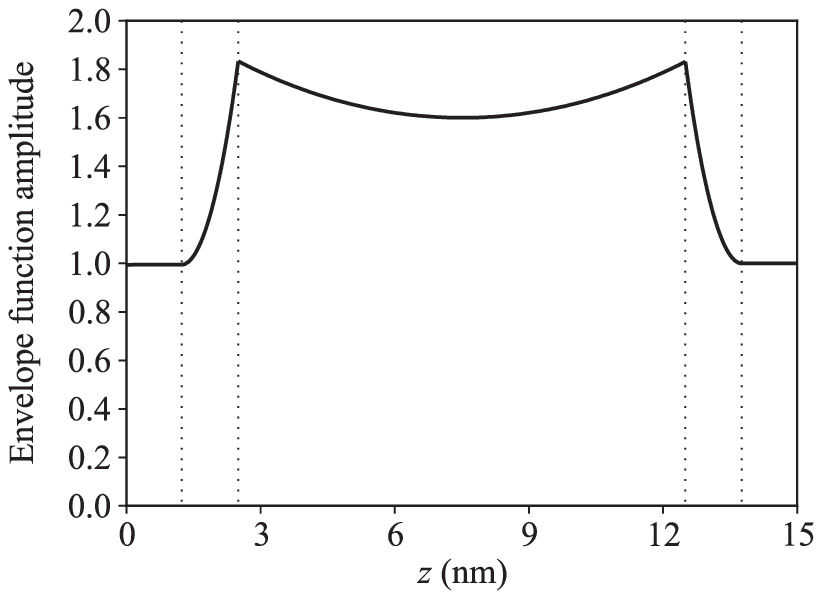}
\caption{\label{Fig6} Amplitude of the envelope function (12) in the structure of Fig. 4 for the energy of maximum transmittance.}
\end{figure}

\section{Conclusions}
Exploiting the formal analogy of electromagnetic wave equation and Schr{\"o}dinger equation we have transfered the idea of electromagnetic perfect tunneling into the semiconductor domain. Using the well accepted Kane model of semiconductor we have particularly shown that the perfect tunneling can be found in 1D semiconductor heterostructures composed of HgCdTe ternary alloys exhibiting all the features of the electromagnetic phenomenon. We feel that the reported results can excite more interest for extending the physical concept of metamaterials into the semiconductor domain leading to new interesting applications.

\begin{acknowledgments}
This work has been supported by the Spanish Ministry of Science and Innovation and European Union FEDER funds (project No. CSD2008-00066), by the Czech Grant Agency (project No. 102/09/0314), by the Grant Agency of the Czech Academy of Sciences (project No. KAN400100652), by the Czech Technical University in Prague (project No. SGS10/271/OHK3/3T/13), and by The Ministry of Education of Czech Republic (project No. MSM 6840770014). Authors are also very grateful to Raul Rodriguez Berral from the department of Applied Physics I at University of Seville for many valuable discussions.
\end{acknowledgments}

\end{document}